\documentclass{llncs}
\usepackage{lscape,lingmacros,prooftree,qtree,amssymb,stmaryrd,txfonts}
\usepackage{stmaryrd}
\usepackage{graphicx}
\usepackage{latexsym}
\usepackage{graphicx}
\title{The Hidden Structural Rules of the Discontinuous Lambek Calculus}
\author{Oriol Valent\'{\i}n}
\newcommand{\disp}[1]{\enumsentence{#1}}

\newcommand{\techterm}[1]{{\it #1}}
\newcommand{\AL}{\mbox{\bf L}}

\newcommand{\wrapk}{\mbox{$\times_k$}}

\newcommand{\mycomment}[1]{}
\newcommand{\product}{\mbox{$\bullet$}}

\newcommand{\bsl}{\mbox{$\backslash$}}
\newcommand{\yields}{\mbox{\ $\Rightarrow$\ }}
\newcommand{\oneyields}{\mbox{\ $\rightarrow$\ }}

\newcommand{\tb}{\hspace*{0.25in}}

\newcommand{\oscott}{\mbox{$\llbracket$}}
\newcommand{\cscott}{\mbox{$\rrbracket$}}

\newcommand{\infix}{\mbox{$\downarrow$}}
\newcommand{\extract}{\mbox{$\uparrow$}}
\newcommand{\dprod}{\mbox{$\odot$}}
\newcommand{\seg}[2]{\mbox{$\sqrt[#1]{#2}$}}
\newcommand{\segfst}[1]{\mbox{$\sqrt[0]{#1}$}}
\newcommand{\segsnd}[1]{\mbox{$\sqrt[1]{#1}$}}

\newcommand{\bydef}{\mbox{$\stackrel{def}{=}$}}
\newcommand{\sep}{\mbox{$[]$}}
\newcommand{\prf}[1]{\noindent{\bf Proof}. #1 $\square$}
\newcommand{\vect}[1]{\overrightarrow{#1}}

\newcommand{\add}{\mbox{$+$}}
\newcommand{\nil}{\mbox{$0$}}
\newcommand{\one}{\mbox{$1$}}

\newcommand{\D}{\mbox{$\mathbf D$}}

\newcommand{\vtab}{\ \vspace{2.5ex}\\}

\newcommand{\defn}[2]{\enumsentence{
{\bf Definition} ({\em #1})\\\ \\#2}}
\newcommand{\lema}[2]{\enumsentence{
{\bf Lemma} ({\em #1})\\\ \\#2}}
\newcommand{\lemawt}[1]{\enumsentence{
{\bf Lemma} \\#1}}% Lemma without title
\newcommand{\rmark}[1]{\enumsentence{
{\bf Remark} \\#1}}% Remark
\newcommand{\examplewt}[1]{\enumsentence{
{\bf Example} \\#1}}% Example without title

\newcommand{\thm}[2]{\enumsentence{
{\bf Theorem} ({\em #1})\\\ \\#2}}

\newcommand{\comp}[1]{\mbox{$\circ_{#1}$}}%Composition: basic mode concatenation. Otherwise, wrapping at the i-th position%
\newcommand{\equi}{\mbox{$\sim$}}% OLD: \newcommand{\equi}{\mbox{$\leftrightarrow^*$}}
\newcommand{\myields}{\mbox{\ $\rightarrow$\ }}
\newcommand{\structterm}{\mbox{$\mathbf{StructTerm}$}}

\newcommand{\Eqd}{\mbox{$\mathbf{Eq_D}$}}

\institute{Universitat Polit\`ecnica de Catalunya
}
\begin{document}
\maketitle
\begin{abstract}
The sequent calculus $\mathbf{sL}$ for the Lambek calculus \AL{} (\cite{lambek:mathematics}) has no structural rules.
Interestingly, $\mathbf{sL}$ is equivalent to a multimodal calculus $\mathbf{mL}$, which consists of the nonassociative Lambek calculus with the structural rule of associativity. This paper proves that the sequent calculus or \techterm{hypersequent} calculus $\mathbf{hD}$ of the discontinuous Lambek calculus\footnote{In \cite{mvf:2009} and \cite{mvf:tdc}, the term \techterm{displacement calculus} is used instead of Discontinous Lambek Calculus as in \cite{mfv:iwcs07} and \cite{valentin:phd}.} (\cite{mfv:iwcs07}, \cite{mv:la} and \cite{mvf:tdc}), which like $\mathbf{sL}$ has no structural rules, is also equivalent to an $\omega$-sorted multimodal calculus $\mathbf{mD}$. More concretely, we present a faithful embedding translation $(\cdot)^\sharp$ between $\mathbf{mD}$ and $\mathbf{hD}$ in such a way that it can be said that $\mathbf{hD}$
absorbs the structural rules of $\mathbf{mD}$.
\end{abstract}

% Methaphor
%\section{A Continuous Metaphor: $\mathbf{NL+Assc}$ and $\mathbf{L}$}

\section{The Discontinuous Lambek Calculus {\D} and its Hypersequent Syntax}
% insertion from proof theory Chapter dlc_thesis
{\D} is model-theoretically motivated, and the key to its conception is the class
$\mathbf{FreeDisp}$ of displacement algebras. We need some definitions:

\defn{Syntactical Algebra}{A
\techterm{syntactical algebra\/}
is a free algebra $(L, \add, \nil, \one)$ of arity $(2, 0, 0)$ such that
$(L, \add, \nil)$ is a monoid and \one{} is a prime. I.e.\
$L$ is a set, $\nil\in L$ and \add{} is a binary operation on $L$
such that for all $s_1, s_2, s_3, s\in L$,
$$
\begin{array}{rcll}
s_1\add(s_2\add s_3) & = & (s_1\add s_2)\add s_3 & \mbox{associativity}\\
\nil\add s = & s & = s\add\nil & \mbox{identity}
\end{array}$$
}
The distinguished constant \one{} is called a {\em separator}.

\defn{Sorts}{The {\em sorts\/} of
discontinuous Lambek calculus are the naturals $0, 1, \ldots$.
The sort $S(s)$ of an element $s$ of a syntactical algebra $(L, \add, \nil, \one)$ is defined by
the morphism of monoids $S$ to the additive monoid of
naturals defined thus:
$$
\begin{array}{rcll}
S(\one) & = & 1\\
S(a) & = & 0 & \mbox{for a prime $a\neq1$}\\
S(s_1\add s_2) & = & S(s_1)+S(s_2)
\end{array}
$$}
I.e.\ the sort of a syntactical element is simply the number
of separators it contains;
we require the separator \one{} to be a prime
and the syntactical algebra to be free in order to ensure that
this induction is well-defined.
\defn{Sort Domains}{Where $(L, \add, \nil,\one)$
is a syntactical algebra,
the {\em sort domains\/} $L_i$ of sort $i$ of generalized
discontinuous Lambek calculus are defined as follows:
$$
\begin{array}{rcl}
L_i & = & \{s| S(s) = i\}, i\ge 0
\end{array}
$$}

\defn{Displacement Algebra}{
The \emph{displacement algebra} defined by a syntactical algebra $(L, \add, 0, \one)$
is the $\omega$-sorted algebra with the $\omega$-sorted signature $\Sigma_D=(\oplus, \{\otimes_{i+1}\}_{i\in\omega},0,1)$ with sort functionality $((i,j\rightarrow i+j)_{i,j\in\omega},(i+1,j\rightarrow i+j)_{i,j\in\omega},0,1)$:

$$(\{L_i\}_{i\in\omega},\add,
\{\times_{i+1}\}_{i\in\omega}, 0, 1)$$

where:

$$
\begin{array}{|l||l|}
\hline
\mbox{operation} & \mbox{is such that}\\
\hline\hline
\add: L_i\times L_j\rightarrow L_{i+j} & \begin{minipage}{43ex}as in the syntactical
algebra\end{minipage}\\
\hline
\times_k: L_{i+1}\times L_j\rightarrow L_{i+j} & \begin{minipage}{43ex}$\times_k(s, t)$ is the result of
replacing the $k$-th separator in $s$ by $t$\end{minipage}\\
\hline
\hline
\end{array}$$
}\label{chapt4opsDA}

% $$
% \begin{array}{|l|}
% \hline
% a\rightarrow b\\
% \hline
% \end{array}
% $$

\noindent The sorted types of the discontinuous Lambek Calculus, $\mathbf{D}$, which we will define residuating with respect 
to the sorted operations in (\ref{chapt4opsDA}), are defined by mutual recursion in Figure~\ref{chapt4typesD}. $\mathbf{D}$ types are to be interpreted as subsets of $L$ and satisfy what we call 
the \techterm{principle of well-sorted inhabitation}:

\begin{figure}
$$\begin{array}{rcll}
{\cal F}_i & ::= & {\mathcal{A}}_i &\mbox{where }{\mathcal{A}}_i\mbox{ is the set of atomic types of sort }i\\\\
{\cal F}_{0}& ::= & I&\mbox{Continuous unit}\\
{\cal F}_{1}& ::= & J&\mbox{Discontinuous unit}\\\\ 
{\cal F}_{i+j} & ::= & {\cal F}_i\product{\cal F}_j&\mbox{continuous product}\\
{\cal F}_j & ::= & {\cal F}_i\bsl{\cal F}_{i+j}&\mbox{continuous under}\\
{\cal F}_i & ::= & {\cal F}_{i+j}/{\cal F}_j&\mbox{continuous over}\\\\
{\cal F}_{i+j} & ::= & {\cal F}_{i+1}\dprod_k{\cal F}_j&\mbox{discontinuous product}\\
{\cal F}_j & ::= & {\cal F}_{i+1}\infix_k{\cal F}_{i+j}&\mbox{discontinuous extract}\\
{\cal F}_{i+1} & ::= & {\cal F}_{i+j}\extract_k{\cal F}_j&\mbox{discontinuous infix}
 \end{array}$$
\caption{The sorted types of $\mathbf{D}$}
\label{chapt4typesD}
\end{figure}

\disp{
$
\begin{array}{|l|}
\hline
\mbox{\textbf{Principle of well-sorted inhabitation:}}\\
\mbox{If }A \mbox{ is a type of sort }i,\mbox{ }\oscott A\cscott\subseteq L_i\\
\hline
\end{array}
$}\label{chapt4princwellsortedinh}

\noindent Where $\oscott\cdot\cscott$ is the syntactical interpretation in a given displacement algebra w.r.t.\ a valuation $v$. I.e.\ every syntactical inhabitant of $\oscott A\cscott$ has the same sort. The connectives and their syntactical interpretations are shown in Figures~\ref{chapt4typesD} and \ref{typeint}.
This syntactical interpretation is called the \techterm{standard syntactical interpretation}.
Given the functionalities of the operations
with respect to which the connectives are defined,
the grammar defining by
mutual recursion the sets ${\cal F}_i$ of types of sort $i$ on the basis of sets
${\cal A}_i$ of atomic types, and the homomorphic \emph{syntactical sort map} $S$ sending
types to their sorts, are as shown in Figure~\ref{typesort}.
When $A$ is an arbitrary type, we will frequently write in latin lower-case the type in order to
refer to its sort $S(A)$, i.e.:

$$a\,\bydef\,S(A)$$
The syntactical sort map is to syntax what the semantic type
map is to semantics: both homomorphisms mapping syntactic
types to the datatypes of the respective components of their
inhabiting signs in the
dimensions of language in extension: form/signifier
and meaning/signified.
%\footnote{If we had continuous and discontinuous product units
%$I=\{\nil\}$ and $J=\{\one\}$ we could define away all the unary connectives by just the two
%nullary connectives as follows:
%$$
%\begin{array}{rclrclrclrcl}
%\leftproj A & = & A/J & \leftinj A & = & A\product J &
%\rightproj A & = & J\bsl A & \rightinj A & = & J\product A\\
%\splitk{k}A & = & A\extract_k I & \bridgek{k}A & = & A\dprod_k I &
%\split A & = & A\extract I & \bridge A & = & A\dprod I
%\end{array}
%$$
%There are four reasons why we do not go down this path: 1) we would require a
%dummy semantics for the units, 2) lexical assignments to units can challenge
%decidability of recognition, 3) proof nets for units are problematic,
%and 4) there would appear to be problems of incompleteness.}

\begin{figure}
\begin{small}
$
\begin{array}[t]{rcll}
\oscott I\cscott & = &\{0\}&\mbox{ continuous unit}\\
\oscott J\cscott & = &\{1\}&\mbox{ discontinuous unit}\\
\oscott A \cscott & \subseteq &L_i\mbox{ for some } i\in\omega &A\in{\mathcal{A}}_i\\\\
\oscott A\product B\cscott & = &
\{s_1\add s_2|\ s_1\in\oscott A\cscott\ \&\
s_2\in\oscott B\cscott\} & \mbox{(continuous) product}\\
\oscott A\bsl C\cscott & = & \{s_2|\ \forall s_1\in\oscott A\cscott,
s_1\add s_2\in\oscott C\cscott\} & \mbox{under}\\
\oscott C/B\cscott & = & \{s_1|\ \forall s_2\in\oscott B\cscott,
s_1\add s_2\in\oscott C\cscott\} & \mbox{over}\\\\
\oscott A\dprod_k B\cscott & = &
\{\times_k(s_1, s_2)|\ s_1\in\oscott A\cscott\ \&\
s_2\in\oscott B\cscott\} & k>0
 \mbox{ \normalfont{deterministic discontinuous product}}\\
\oscott A\infix_k C\cscott & = & \{s_2|\ \forall s_1\in\oscott A\cscott,
\times_k(s_1, s_2)\in\oscott C\cscott\} & k>0\mbox{ \normalfont{deterministic discontinuous infix}}\\
\oscott C\extract_k B\cscott & = & \{s_1|\ \forall s_2\in\oscott B\cscott,
\times_k(s_1, s_2)\in\oscott C\cscott\} & k>0\mbox{ \normalfont{deterministic discontinuous extract}}\\
\\\\
\end{array}
$
\end{small}
\caption{Standard syntactical interpretation of $\mathbf{D}$ types}
\label{typeint}
\end{figure}

\begin{figure}
$$\begin{array}{rclrcll}
{\cal F}_i & ::= & {\mathcal{A}}_i & S(A) & = & i & \mbox{for $A\in{\cal A}_i$}\\\\
{\cal F}_{0}& ::= & I& S(I) & = & 0 &\\
{\cal F}_{1}& ::= & J& S(J) & = & 1 &\\\\  
{\cal F}_{i+j} & ::= & {\cal F}_i\product{\cal F}_j & S(A\product B) & = & S(A)+S(B)\\
{\cal F}_j & ::= & {\cal F}_i\bsl{\cal F}_{i+j} & S(A\bsl C) & = & S(C)-S(A)\\
{\cal F}_i & ::= & {\cal F}_{i+j}/{\cal F}_j & S(C/B) & = & S(C)-S(B)\\\\
{\cal F}_{i+j} & ::= & {\cal F}_{i+1}\dprod_k{\cal F}_j & S(A\dprod_k B) & = & S(A)+S(B)-1
 & 1\le k\le i+1\\
{\cal F}_j & ::= & {\cal F}_{i+1}\infix_k{\cal F}_{i+j} & S(A\infix_k C) & = & S(C)+1-S(A)
 & 1\le k\le i+1\\
{\cal F}_{i+1} & ::= & {\cal F}_{i+j}\extract_k{\cal F}_j & S(C\extract_k B) & = & S(C)+1-S(B)
 & 1\le k\le i+1
 \end{array}$$
\caption{Sorted $\mathbf{D}$ types, and syntactical sort map for $\mathbf{D}$}
\label{typesort}
\end{figure}

Observe also that (modulo sorting)
$(\bsl, \product, /; \subseteq)$ and
$(\infix_k, \dprod_k, \extract_k; \subseteq)$ are residuated triples:
\disp{
$
\begin{array}[t]{rcccl}
\oscott B\cscott\subseteq \oscott A\bsl C\cscott & \mbox{iff} & \oscott A\product B\cscott\subseteq \oscott C\cscott & \mbox{iff} & \oscott A\cscott\subseteq \oscott C/B\cscott\\
\oscott B\cscott\subseteq \oscott A\infix_k C\cscott & \mbox{iff} & \oscott A\dprod_k B\cscott\subseteq \oscott C\cscott & \mbox{iff} & \oscott A\cscott\subseteq \oscott C\extract_k B\cscott\\
\end{array}
$}

% end insertion

% Hyperconfiguration and vectorial notation

The types of \D{} are sorted into types ${\cal F}_i$ of sort $i$ 
interpreted as sets of strings of sort $i$ as shown in Figure~\ref{typesandint}
where $k\in\omega^+$.
\begin{figure}
$$\footnotesize
\begin{array}{rclrcll}
{\cal F}_j & := & {\cal F}_i\bsl{\cal F}_{i{+}j} &
[A\bsl C] & = & \{s_2|\ \forall s_1\in[A], s_1\add s_2\in[C]\} & \mbox{under}\\
{\cal F}_i & := & {\cal F}_{i{+}j}/{\cal F}_j &
[C/B] & = & \{s_1|\ \forall s_2\in[B], s_1\add s_2\in[C]\} & \mbox{over}\\
{\cal F}_{i{+}j} & := & {\cal F}_i\product{\cal F}_j &
[A\product B] & = & \{s_1\add s_2|\ s_1\in[A]\ \&\ s_2\in[B]\} & \mbox{product}\\
{\cal F}_0 & := & I & [I] & = & \{\nil\} & \mbox{product unit}\\
{\cal F}_j & := & {\cal F}_{i{+}1}\infix_k{\cal F}_{i{+}j} &
[A\infix_k C] & = & \{s_2|\ \forall s_1\in[A], s_1\wrapk s_2\in[C]\} & \mbox{infix}\\
{\cal F}_{i{+}1} & := & {\cal F}_{i{+}j}\extract_k{\cal F}_j &
[C\extract_k B] & = & \{s_1|\ \forall s_2\in[B], s_1\wrapk s_2\in[C]\} & \mbox{extract}\\
{\cal F}_{i{+}j} & := & {\cal F}_{i{+}1}\dprod_k{\cal F}_j &
[A\dprod_k B] & = & \{s_1\wrapk s_2|\ s_1\in[A]\ \&\ s_2\in[B]\} & \mbox{disc.\ product}\\
{\cal F}_1 & := & J & [J] & = & \{\one\} & \mbox{disc.\ prod.\ unit}\\
\end{array}
$$
\caption{Types of the Discontinuous Lambek Calculus {\D} and their interpretation}
\label{typesandint}
\end{figure}

If one wants to absorb the structural rules of a Gentzen sequent system in a 
substructural logic, one has to discover a convenient data structure for the antecedent 
and the succedent of sequents. We will now consider the \techterm{Hypersequent syntax}\footnote{Term which must not be confused with Avron's
hypersequents (\cite{avron:91}).} from \cite{mfv:iwcs07}. The reason for using the prefix \techterm{hyper} in the term
\techterm{sequent} is that the data-structure proposed 
is quite nonstandard. 
We define now  what we call the set of types segments:
\defn{Type Segments}{In hypersequent
calculus we define
the {\em types segments\/}
${\cal SF}_k$
of sort $k$:

$$
\begin{array}{rcll}
{\cal SF}_0 & ::= & A & \mbox{for\ } A\in{\mathcal{F}}_0\\
{\cal SF}_a & ::= & \seg{i}{A} & \mbox{ for }A\in{\mathcal{F}}_a\mbox{ and }0\leq i\leq a=S(A)
\end{array}
$$}

\noindent Types segments of sort $0$ are types. But, types segments of sort greater than
$0$ are no longer types. Strings of types segments can form meaningful logical material like
the set of hyperconfigurations, which we now define. The {\em hyperconfigurations\/}
${\cal O}$ are defined unambiguously by
mutual recursion as follows, where $\Lambda$ is the empty string and \sep{} is the metalinguistic separator::
$$
\begin{array}{rcl}
{\cal O} & ::= & \Lambda \\%& A\ \mbox{for\ } S(A)=0\\
%{\cal O} & ::= & \sep\\
{\cal O} & ::= & A, {\cal O}\ \mbox{for\ } S(A)=0\\
{\cal O} & ::= & \sep, {\cal O}\\
{\cal O} & ::= & \segfst{A},
{\cal O}, \segsnd{A}, \ldots, \seg{a-1}{A}, {\cal O},
\seg{a}{A}, {\cal O}\\
 && \mbox{for\ } a=S(A)>0
\end{array}$$
\noindent The syntactical interpretation of $\seg{0}{A},{\cal O},\seg{1}{A},{\cal O},\ldots,\hspace{-0.3cm}\seg{a-1}{A},{\cal O},\seg{a}{A}$ consists of syntactical elements $\alpha_0\add\beta_1\add\alpha_1\add\cdots\add$ $\alpha_{n{-}1}\add\beta_n\add\alpha_n$
where $\alpha_0\add\one\add\alpha_1\add\cdots\add\alpha_{n{-}1}\add\one\add\alpha_n\! \in \oscott A\cscott$ and
 $\beta_1\in \oscott\Delta_1\cscott, \ldots, \beta_n \in\oscott\Delta_n\cscott$.
The syntax in which set ${\cal O}$ has been defined, is called \techterm{string-based hypersequent syntax}. An equivalent syntax for ${\cal O}$ is called \techterm{tree-based hypersequent syntax}
which was defined in \cite{mv:la}, \cite{mvf:tdc}.

\noindent In string-based notation the \techterm{figure} $\vect{A}$ of a type $A$ is defined as follows:
\disp{$
\vect{A} = \left\{
\begin{array}{ll}
A & \mbox{if\ } s(A)=0\\
\seg{0}{A},\sep,\seg{1}{A},\sep,\ldots,\hspace{-0.3cm}\seg{a-1}{A},\sep,\seg{a}{A} & \mbox{if\ } s(A)>0
\end{array}\right.$
}

The sort of a hyperconfiguration is the number of metalinguistic separators it contains.
Where $\Gamma$ and $\Phi$ are hyperconfigurations and the sort of $\Gamma$ is
at least $1$,
$\Gamma|_k\Phi$ ($k\in \omega^+$) signifies the hyperconfiguration which is the result of
replacing the k-th separator in $\Gamma$ by $\Phi$.
Where  $\Gamma$ is a hyperconfiguration of sort $i$ and $\Phi_1, \ldots, \Phi_i$ are hyperconfigurations,
the \techterm{generalized wrap} $\Gamma\otimes\langle \Phi_1, \ldots, \Phi_i\rangle$ is the
result of simultaneously replacing the successive separators in $\Gamma$ by $\Phi_1, \ldots, \Phi_i$
respectively.
$\Delta\langle\Gamma\rangle$
abbreviates
$\Delta(\Gamma\otimes\langle \Delta_1, \ldots, \Delta_i\rangle)$.

A \techterm{hypersequent\/} $\Gamma\yields A$ comprises an  
antecedent hyperconfiguration in string-based notation of sort $i$ and a succedent type $A$ of sort $i$.
The hypersequent calculus for \D{} is as shown in Figure~\ref{Dseqcalc}
where $k\in\omega^+$. 
Like {\AL}, $\mathbf{hD}$ has no structural rules. 
\begin{figure}
\begin{center}
$\prooftree
\justifies
\vect{A}\yields A
\using id
\endprooftree \tb
\prooftree
\Gamma\yields A \tb
\Delta\langle \vect{A}\rangle\yields B
\justifies
\Delta\langle\Gamma\rangle\yields B
\using Cut
\endprooftree$
\vtab
$\prooftree
\Gamma\yields A \tb
\Delta\langle\vect{C}\rangle\yields D
\justifies
\Delta\langle\Gamma, \vect{A\bsl C}\rangle\yields D
\using \bsl L
\endprooftree \tb
\prooftree
\vect{A}, \Gamma\yields C
\justifies
\Gamma\yields A\bsl C
\using \bsl R
\endprooftree$
\vtab
$\prooftree
\Gamma\yields B \tb
\Delta\langle\vect{C}\rangle\yields D
\justifies
\Delta\langle\vect{C/B}, \Gamma\rangle\yields D
\using / L
\endprooftree \tb
\prooftree
\Gamma, \vect{B}\yields C
\justifies
\Gamma\yields C/B
\using / R
\endprooftree$
\vtab
$\prooftree
\Delta\langle\vect{A}, \vect{B}\rangle\yields D
\justifies
\Delta\langle\vect{A\product B}\rangle\yields D
\using \product L
\endprooftree \tb
\prooftree
\Gamma_1\yields A\tb\Gamma_2\yields B
\justifies
\Gamma_1, \Gamma_2\yields A\product B
\using \product R
\endprooftree
$
\vtab
$\prooftree
\Delta\langle\Lambda\rangle\yields A
\justifies
\Delta\langle\vect{I}\rangle\yields A
\using IL
\endprooftree\tb
\prooftree
\justifies
\Lambda\yields I
\using IR
\endprooftree
$
\vtab
$\prooftree
\Gamma\yields A \tb
\Delta\langle\vect{C}\rangle\yields D
\justifies
\Delta\langle\Gamma|_k\vect{A\infix_k C}\rangle\yields D
\using \infix_k L
\endprooftree \tb
\prooftree
\vect{A}|_k\Gamma\yields C
\justifies
\Gamma\yields A\infix_k C
\using \infix_k R
\endprooftree$
\vtab
$\prooftree
\Gamma\yields B \tb
\Delta\langle\vect{C}\rangle\yields D
\justifies
\Delta\langle\vect{C\extract_k B}|_k\Gamma\rangle\yields D
\using \extract_k L
\endprooftree \tb
\prooftree
\Gamma|_k\vect{B}\yields C
\justifies
\Gamma\yields C\extract_k B
\using \extract_k R
\endprooftree$
\vtab
$\prooftree
\Delta\langle\vect{A}|_k\vect{B}\rangle\yields D
\justifies
\Delta\langle\vect{A\dprod_k B}\rangle\yields D
\using \dprod_k L
\endprooftree \tb
\prooftree
\Gamma_1\yields A\tb\Gamma_2\yields B
\justifies
\Gamma_1|_k\Gamma_2\yields A\dprod_k B
\using \dprod_k R
\endprooftree$
\vtab
$\prooftree
\Delta\langle\sep\rangle\yields A
\justifies
\Delta\langle\vect{J}\rangle\yields A
\using JL
\endprooftree\tb
\prooftree
\justifies
\sep\yields J
\using JR
\endprooftree
$\end{center}
\caption{Hypersequent calculus $\mathbf{hD}$}
\label{Dseqcalc}
\end{figure}

Morrill and Valent\'{\i}n (2010)\cite{mv:la} proves Cut-elimination for the $k$-ary discontinuous Lambek
calculus, $k>0$.
As a consequence {\D}, like {\AL}, enjoys in addition
the subformula property, decidability, and the finite reading property.

% absorbing the structural rules
\section{$\mathbf{hD}$: Absorbing the Structural Rules of a Sorted Multimodal Calculus}
We consider now a sorted multimodal calculus $\mathbf{mD}$ with a set of structural rules
$\mathbf{Eq_D}$ we present in the following lines. Figure~\ref{chapt4mDIfig} shows the logical rules
of $\mathbf{mD}$ and Figure~\ref{chapt4mDIfigI} shows the structural rules \Eqd{} integrated in $\mathbf{mD}$. 
This sequent calculus is non
standard in two senses. Types and structural
terms are sorted. Moreover, there are two structural term constants which stand
respectively for the continuous unit and discontinuous unit.
Structural term constructors are of two kinds: $\comp{}$ (which
stands for term concatenation) and $\comp i$ (which stands for term
wrapping at the i-th position, $i\in\omega^+$). Again, as in the case of
sorted types, structural terms are defined by mutual recursion and
the \techterm{sort map} is computed in a similar way (see
(\ref{chapt4structterm})).

$X[Y]$ denotes a structural term with a distinguished position
occupied by the structural term Y. If A, X are respectively a type
and a structural term, then \emph{a} and \emph{x} denote their
sorts. We are interested in the cardinality of the set $\mathcal{F}$ of types of $\mathbf{D}$ 
and their structure. Consider the following lemma:

\lemawt{The set of types $\mathcal{F}$ is countably infinite iff the set of atomic types is countable. Moreover we have that:
$$
\begin{array}{lll}
\mathcal{F}&=&\displaystyle\bigcup_{i\in\omega}{\mathcal{F}}_{i}\\
{\mathcal{F}}_{i}&=& (A_{ij})_{j\in\omega}
\end{array}
$$}
\prf{The proof can be carried out by coding in a finite alphabet the set of types $\mathcal{F}$.
Of course, it is crucial that the set of sorted atomic types forms a denumerable set.}\\% end of proof

Let $\mathbf{StructTerm}_{\mbox{D}}[\mathcal{F}]$ be the $\omega$-sorted algebra
over the signature $\Sigma_D=(\{\comp{}\}\cup (\comp{i+1})_{i\in\omega},\mathbb{I},\mathbb{J})$.
The sort functionality of $\Sigma_D$ is:

$$((i,j\rightarrow i+j)_{i,j\in\omega},(i+1,j\rightarrow i+j)_{i,j\in\omega},0,1)$$

\noindent Observe that the operations $\comp{}$ and $\comp{i}$'s (with $i>0$) are sort polymorphic.
In the following, we will abbreviate $\mathbf{StructTerm}_{\mbox{D}}[\mathcal{F}]$ by $\mathbf{StructTerm}$.
The set of \techterm{structural terms} can be defined in BNF notation as follows:

%%%%%%%%%%%%%%%%%%%%%%%%%%%%%%%%%%%%%%%%%%%%%%%%%%%%%%%%%%%%%%%%%
% Sorted structural terms for mD
% Bijection with T_{\Sigma_D}[X]
%%%%%%%%%%%%%%%%%%%%%%%%%%%%%%%%%%%%%%%%%%%%%%%%%%%%%%%%%%%%%%%%%

\disp{
$\begin{array}{rclrcll}
\structterm_0 & ::= & \mathbb{I}\\
\structterm_1 & ::= & \mathbb{J}\\
\structterm_i & ::= & {\cal F}_i\\
\structterm_{i+j} & ::= & \structterm_i\comp{}\structterm_j \\
\structterm_{i+j} & ::= & \structterm_{i+1}\comp{k}\structterm_j
 \end{array}$}\label{chapt4structterm}

\noindent It is clear that the sort of $\structterm_i$ and the collections of set $(A_{ij})_{j\in\omega}$ ($i\in \omega$) are such that:
$$
\begin{array}{lll}
S(\structterm_i) &=& i\\
S(A_{ij})&=&i
\end{array}
$$
\noindent We realize that $\mathbf{StructTerm}$ looks like an $\omega$-sorted term algebra.
This intuition is correct for the $\omega$-graduated set $\mathcal{F}$ with the collections $(A_{ij})_{j\in\omega}$
plays the role of an $\omega$-graduated set of a variables of an
$\omega$-sorted term algebra $T_{\Sigma_D}[X]$ with signature $\Sigma_D$.

% Let us consider the following bijective mapping $f$ from the set of variables
%$X$ of Chapter 3 into the set of types $\mathcal{F}$.\footnote{The existence of this bijection is not difficult to see 
%given that the set of atomic variables and atomic types are denumerable.}
%
%$$
%\begin{array}{llll}
%f:& X&\longrightarrow&\mathcal{F}\\
%& x_{ij}&\mapsto& A_{ij}\\
%\end{array}
%$$ 
%
%
%\noindent This bijection is such that (for $i,j\in\omega$):
% 
%
%$$S(x_{ij}) = S(f(x_{ij})) = S(A_{ij})=i$$
%
%\noindent So, for every $i\in\omega$, the sets $(x_{ij})_{j\in\omega}$ and $(A_{ij})_{j\in\omega}$ 
%are respectively the set of sorted variables of sort $i$ and the set of types of sort $i$. $f$ extends recursively
% to $f^*$ as follows:
%
%\disp{
%$$
%\begin{array}{llll}
%f^*:& T_{\Sigma_D}[X]&\longrightarrow&\mathbf{StructTerm}\\
%& 0&\mapsto& \mathbb{I}\\
%& 1&\mapsto& \mathbb{J}\\
%& x_{ij}&\mapsto& A_{ij}\\
%& t\oplus s&\mapsto& f(t)\comp{}f(s)\\
%& t\otimes_i s&\mapsto& f(t)\comp{i}f(s)
%\end{array}
%$$}\label{chapt4mapf}
%
%\noindent Since $f$ is bijective and $f$ extends recursively to $f^*$, it is easy to prove by induction
%on the structure of $\mathbf{StructTerm}$ that $f^*$ is bijective. Notice that in fact $f^*$ is a sorted $\Sigma_D$-isomorphism.
%% Correspondance between T_{\Sigma_D}[X] and \mathbf{StructTerm}

%\subsubsection{}
We need to define some important relations between structural terms.

\defn{Wrapping and Permutable Terms}{
Given the term $(T_1\circ_i T_2)\circ_j T_3$, we say that:
\begin{itemize}
  \item[(P1)] $T_2\prec_{T_1}T_3$ iff $i+t_2-1<j$.
  \item[(P2)] $T_3\prec_{T_1}T_2$ iff $j<i$.
  \item[(O)] $T_2 \between_{T_1} T_3$ iff $i\leq j\leq i+t_2-1$.
\end{itemize}
}\label{wrappermtermdef}% end definition

\noindent  Observe that in a term like $(T_1\circ_i T_2)\circ_j T_3$, if $(P1)$
or $(P2)$ hold, $(O)$ does not apply. Conversely, if $(O)$ is
applicable, neither $(P1)$ nor $(P2)$ hold. If $T_2\prec_{T_1}T_3$
(respectively $T_3\prec_{T_1}T_2$), we say that $T_2$ and $T_3$
(respectively $T_3$ and $T_2$) \techterm{permute} in $T_1$. Otherwise,
if $(O)$ holds, we say that $T_2$ \techterm{wraps} $T_3$ in $T_1$. 
\examplewt{ 
Supose that $T_1=A$ where $A$ is an arbitrary type of sort $3$, and $T_2,T_3$ are arbitrary structural terms.
Let $a_0 + 1 + a_1 + 1 + a_2 + 1 + a_3$ be an element of $\oscott A\cscott$ in a displacement model $\mathcal{M}$.
Suppose $S(T_2)=3$. Consider now:
$$(A\comp{2}T_2)\comp{5} T_3$$
\noindent According to definition (\ref{wrappermtermdef}), $T_2\prec_A T_3$, for $2+S(T_2)-1=4 < 5$. The intuition of this relation is the following. 
Interpreting in $M$ we have that:
\disp{$\oscott (A\comp{2}T_2)\comp{5} T_3\cscott= a_0+1+a_1+\oscott T_2\cscott+a_2+\oscott T_3\cscott+a_3$}\label{example1}

We clearly see that the string $\oscott T_2\cscott$ precedes the occurrence of $\oscott T_3\cscott$. Similarly, if we have $T_3\prec_{A} T_2$ in
$(A\comp{i} T_2)\comp{j} T_3$, the occurrence of $\oscott T_3\cscott$ precedes $\oscott T_2\cscott$. Finally, if $T_2\between_{A} T_3$ then
$\oscott T_2\cscott$ wraps $\oscott T_3\cscott$, i.e.\ $\oscott T_3\cscott$ is intercalated in $\oscott T_2\cscott$.
}
% insertion
\noindent We define the following relation between structural terms $\equi$:
\disp{ 
$
T\equi S\mbox{ iff } S\mbox{ is the result of applying one structural rule to a subterm of T}
$
}
\noindent $\equi^*$ is defined to be the reflexive, symmetric and transitive closure of $\equi$.

% end insertion

\begin{figure}
$
A\myields A\mbox{ Id } \tb
\prooftree
S\oneyields A\tb T[A]\oneyields B
\justifies
T[S]\oneyields B
\using Cut
\endprooftree 
$
\vtab
$\prooftree T[\mathbb I]\myields
A\justifies T[I]\myields A\using I L\endprooftree
\tb
\prooftree\justifies \mathbb I\yields I
\using I R
\endprooftree
$
\vtab
$\prooftree T[\mathbb J]\myields
A\justifies T[J]\myields A\using JL\endprooftree
\tb
\prooftree\justifies \mathbb J\yields J
\using J R
\endprooftree
$
\vtab

$\prooftree X\myields A \tb Y[B]\myields C\justifies Y[X\comp{}
A\backslash B]\myields C\using \backslash L\endprooftree 
\tb
\prooftree A\comp{} X\myields B\justifies X\myields A\backslash
B\using \backslash R\endprooftree
$
\vtab
$\prooftree X\myields A \tb Y[B]\myields C\justifies Y[B/A\comp{}
X]\myields C\using / L\endprooftree 
\tb
\prooftree  X\comp{} A\myields B\justifies X\myields B/A
\using / R\endprooftree
$
\vtab
$
\prooftree X\myields A \tb Y[B]\myields C\justifies Y[B\uparrow_i
A\comp{i} X]\myields C\using \uparrow_i L\endprooftree$
\tb
$\prooftree X\comp{i} A\myields B\justifies X\myields
B\uparrow_iA\using \extract_i R\endprooftree$
\vtab
$\prooftree X\myields A \tb Y[B]\myields C\justifies Y[X\comp{i}
A\downarrow_i B]\myields C\using \downarrow_i L\endprooftree$ 
\tb
$\prooftree A\comp{i} X\myields B\justifies X\myields A\downarrow_i
B\using \downarrow_i R\endprooftree$
\vtab
$\prooftree X[A\comp{} B]\myields C\justifies X[A\bullet B]\myields
C\using \bullet L\endprooftree$ 
\tb 
$\prooftree X\myields A \tb
Y\myields B\justifies X\comp{} Y\myields A\bullet B\using \bullet
R\endprooftree$
\vtab
$\prooftree X[A\comp i B]\myields C\justifies X[A\dprod_i B]\myields
C\using \dprod_i L\endprooftree$ 
\tb 
$\prooftree X\myields A \tb
Y\myields B\justifies X\comp i Y\myields A\dprod_i B\using \dprod_i
R\endprooftree$
\caption{The Logical rules of $\mathbf{mD}$}
\label{chapt4mDIfig}
\end{figure}

% Structural rules
\begin{figure}
$\mbox{\textbf{Structural rules for units}}\\$

- Continuous unit:\\

$\prooftree T[X]\myields A\justifies T[\mathbb I\comp{} X]\myields A
\endprooftree\tb \prooftree T[\mathbb I \comp{}X]\myields A\justifies T[X]\myields A
\endprooftree$\tb\prooftree T[X]\myields A\justifies T[X\comp{}\mathbb I]\myields A
\endprooftree \tb \prooftree T[X\comp{}\mathbb I]\myields A\justifies T[X]\myields A
\endprooftree\\

- Discontinuous unit:\\

$\prooftree T[X]\myields A\justifies T[\mathbb J\comp{1} X]\myields A
\endprooftree\tb \prooftree T[\mathbb J \comp{1}X]\myields A\justifies T[X]\myields A
\endprooftree\tb\prooftree T[X]\myields A\justifies T[X\comp{i}\mathbb J]\myields A
\endprooftree \tb \prooftree T[X\comp{i}\mathbb J]\myields A\justifies T[X]\myields A
\endprooftree$\\\\

$\mbox{\textbf{Continuous associativity}}$\\

$\prooftree X[(T_1\comp{}T_2)\comp{} T_3]\myields D \justifies
X[T_1\comp{}(T_2\comp{} T_3)]\myields D\using Assc_c\endprooftree$
\tb $\prooftree X[T_1\comp{}(T_2\comp{} T_3)]\myields D\justifies
X[(T_1\comp{}T_2)\comp{} T_3]\myields D \using
Assc_c\endprooftree$\\\\\vspace{0.5cm}

$\mbox{\textbf{Split-wrap}}\\$

$\prooftree T_1[T_2\comp{}T_3]\myields D\justifies
T_1[(\mathbb{J}\comp{}T_3)\comp{1}T_2]\myields D\using SW\,
\endprooftree$ \tb $\prooftree
T_1[(\mathbb{J}\comp{}T_3)\comp{1}T_2]\myields D\justifies
T_1[T_2\comp{}T_3]\myields D\using SW
\endprooftree$\\\\

$\prooftree T_1[T_2\comp{}T_3]\myields D\justifies
T_1[(T_2\comp{}\mathbb{J})\comp{t_2+1}T_3]\myields D\using
SW\endprooftree$ \tb $\prooftree
T_1[(T_2\comp{}\mathbb{J})\comp{t_2+1}T_3]\myields D \justifies
T_1[T_2\comp{}T_3]\myields D \using SW\endprooftree$
\vtab
$$
\begin{array}{l}
\mbox{\textbf{Discontinuous associativity} $T_2\between_{T_1} T_3$}\\\\
 \prooftree
S[T_1\comp{i}(T_2\comp{j} T_3)]\oneyields C
\justifies
S[(T_1\comp{i} T_2)\comp{i+j-1} T_3)]\oneyields C
\using \mathbf{Assc_d1}
\endprooftree
\tb
\prooftree
S[(T_1\comp{i} T_2)\comp{j} T_3]\oneyields C
\justifies
S[T_1\comp{i}(T_2\comp{j-i+1} T_3)]\oneyields C
\using \mathbf{Assc_d2}
\endprooftree
\end{array}
$$

$$
\begin{array}{l}
\mbox{\textbf{Mixed permutation 1}}\mbox{ case }T_2\prec_{T_1} T_3 \\\\
\prooftree
S[(T_1\comp{i}T_2)\comp{j}T_3]\oneyields C
\justifies
S[(T_1\comp{j-S(T_2)+1}T_3)\comp{i} T_2]\oneyields C
\using \mathbf{MixPerm1}
\endprooftree
\tb
\prooftree
S[(T_1\comp{i} T_3)\comp{j} T_2]\oneyields C
\justifies
S[(T_1\comp{j} T_2)\comp{i+S(T_2)-1}z]\oneyields C
\using
\mathbf{MixPerm1}
\endprooftree
\end{array}
$$

$$
\begin{array}{l}
\mbox{\textbf{Mixed permutation 2}}\mbox{ case }T_3\prec_{T_1} T_2 \\\\
\prooftree
S[(T_1\comp{i}T_2)\comp{j}T_3]\oneyields C
\justifies
S[(T_1\comp{j}T_3)\comp{i+S(T_3)-1} T_2]\oneyields C
\using \mathbf{MixPerm2}
\endprooftree
\tb
\prooftree
S[(T_1\comp{i} T_3)\comp{j} T_2]\oneyields C
\justifies
S[(T_1\comp{j-S(T_3)+1} T_2)\comp{i}T_3]\oneyields C
\using
\mathbf{MixPerm2}
\endprooftree
\end{array}
$$
\caption{Structural Rules of $\mathbf{mD}$}
\label{chapt4mDIfigI}
\end{figure}

\subsection{The Faithful embedding translation $(\cdot)^\sharp$ between $\mathbf{mD}$ and $\mathbf{hD}$}
% (\cdot)^\sharp embedding translation

We consider the following embedding translation from $\mathbf{mD}$ to
$\mathbf{hD}$:
 $$
          \begin{array}{clcr}
               (\cdot)^\sharp:\mathbf{mD} =(\mathcal{F},
\structterm, \myields) & \longrightarrow &
\mathbf{hD}=(\mathcal{F}, \mathcal{O},
  \Rightarrow)     \\
               T\myields  A    & \mapsto   & (T)^\sharp \yields
               (A)^\sharp
          \end{array}
     $$
$(\cdot)^\sharp$ is such that:
$$
      \begin{array}{l}
	A^\sharp= \vect{A}\mbox{ if }A\mbox{ is of sort strictly greater than }0\\
        A^\sharp= A\mbox{ if A is of sort }0\\
        (T_1\circ T_2)^\sharp= T_1^\sharp,T_2^\sharp\\
        (T_1\circ_i T_2)^\sharp= T_1^\sharp |_i T_2^\sharp\\
        \mathbb{I}^\sharp= \Lambda\\
        \mathbb{J}^\sharp= \sep\\
          \end{array}
$$

\noindent\textbf{Collapsing the structural rules}\\\\
Let us see how the structural rules are absorbed in $\mathbf{hD}$. We show here that structural postulates
of $\mathbf{mD}$ collapse into the same textual form when they are mapped through $(\cdot)^\sharp$.
Later we will see that:

 $$\mbox{If } T\equi^* S \mbox{ then } T^\sharp = S^\sharp$$
 
\noindent Moreover will see that for every $A,B,C\in\mathcal{F}$ the following hypersequents are provable in $\mathbf{hD}$:

\disp{
\begin{itemize}
\item \textbf{Continuous associativity}

$$\vect{A\bullet (B\bullet C)}\yields (A\bullet B)\bullet C\mbox{ and }
\vect{(A\bullet B)\bullet C}\yields A\bullet (B\bullet C)$$

\item \textbf{Mixed associativity}
If we have that $B\between_A C$:
$$\vect{A\odot_i(B\odot_j C) } \yields (A\odot_i B)\odot_{i+j-1} C\mbox{ and }
\vect{(A\odot_i B)\odot_{i+j-1} C) } \yields A\odot_i (B\odot_j C)$$

\item \textbf{Mixed permutation}
If we have that $B\prec_A C$:
$$\vect{(A\odot_i B)\odot_{j} C) } \yields (A\odot_{j-b+1} C)\odot_i C \mbox{ and }
\vect{(A\odot_{j-b+1} C)\odot_i C} \yields (A\odot_i B)\odot_{j} C)$$
\noindent If we have that $C\prec_A B$:
$$\vect{(A\odot_i B)\odot_{j} C) } \yields (A\odot_j C)\odot_{i+c-1} C 
\vect{(A\odot_j C)\odot_{i+c-1} C} \yields (A\odot_i B)\odot_{j} C)$$
\item \textbf{Split wrap}:
$$\vect{A\bullet B}\yields (A\bullet J)\odot_{a+1}B\mbox{ and }
\vect{(A\bullet J)\odot_{a+1}B}\yields A\bullet B$$

\noindent and:
$$\vect{(J\bullet B)\odot_1 A}\yields A\bullet B
\mbox{ and }\vect{A\bullet B}\yields (J\bullet B)\odot_1 A$$

\item \textbf{Continuous unit and discontinuous unit}:
$$\vect{A\bullet I}\yields A\mbox{ and } \vect{A}\yields A\bullet I\mbox{ and } 
\vect{I\bullet A}\yields A \mbox{ and }\vect{A}\yields I\bullet A$$

\noindent and:
$$\vect{A\odot_i J}\yields A\mbox{ and } \vect{A}\yields A\odot_i J\mbox{ and } 
\vect{J\odot_1 A}\yields A\mbox{ and } \vect{A}\yields J\odot_1 A$$
\end{itemize}}\label{chapt4disphdthms}

\noindent That $\mathbf{hD}$ absorbs the rules is proved in the following theorem: 

\thm{$\mathbf{hD}$ Absorption of $\Eqd$ Structural Rules}{
For any $T,S\in\mathbf{StructTerm}$, if $T\equi^*S$ then $(T)^\sharp = (S)^\sharp$.
}\label{chapt4hDabsoptionSR}

\prf{
We define a useful notation for vectorial types which will help us to prove the theorem. Where $A$
is an arbitrary type of sort greater than $0$:

\disp{
$
\vect{A}_i^j=\left\{
\begin{array}{l}
\seg{i}{A}\mbox{, if }i=j\\
\vect{A}_i^{j-1},\sep,\seg{j}{A}\mbox{, if }j-i>0
\end{array}
\right.
$}

\noindent Note that $\vect{A}=\vect{A}_0^a$. Now, consider arbitrary types $A,B$ and $C$. 
As usual we denote their sorts respectively by $a$, $b$ and $c$. We have then:
 
\begin{itemize}
 \item Continuous associativity:

$$
\left \{
\begin{array}{lll}
 ((A\comp{} B)\comp{} C)^\sharp &=& (\vect{A},\vect{B}),\vect{C}= \vect{A},\vect{B},\vect{C}\\
  (A\comp{} (B)\comp{} C))^\sharp &=& \vect{A},(\vect{B},\vect{C})= \vect{A},\vect{B},\vect{C}
\end{array}\right.
$$

\item Discontinuous associativity: Suppose that $B\between_A C$\\

\noindent We have that:

$$
\begin{array}{l}
\vect{B}|_j\vect{C}= \vect{B}_0^{i-1},\vect{C},\vect{B}_i^{b}\\
\vect{A}|_i(\vect{B}|_j\vect{C})= \vect{A}_0^{i-1},\vect{B}_0^{j-1},\vect{C},\vect{B}_j^{b},\vect{A}_i^{a}
\end{array}
$$

\noindent On the other hand, we have that:

$$
\begin{array}{l}
\vect{A}|_i\vect{B}= \vect{A}_0^{i-1},\vect{B},\vect{A}_i^{a}
=\vect{A}_0^{i-1},\vect{B}_0^{j-1},\underbrace{\sep}_{(i+j-1)\scriptsize{\mbox{-th }\sep}},\vect{B}_j^{b},\vect{A}_i^{a}\\
\end{array}
$$

\noindent It follows that:

$$(\vect{A}|_i\vect{B})|_{i+j-1}\vect{C}= \vect{A}_0^{i-1},\vect{B}_0^{j-1},\vect{C},\vect{B}_j^{b},\vect{A}_i^{a}$$

\noindent Summarizing:

$$
\left \{
\begin{array}{lll}
 (A\comp{i} (B\comp{j} C))^\sharp &=& \vect{A}_0^{i-1},\vect{B}_0^{j-1},\vect{C},\vect{B}_j^{b},\vect{A}_i^{a}\\
  ((A\comp{i} B)\comp{i+j-1} C)^\sharp &=& \vect{A}_0^{i-1},\vect{B}_0^{j-1},\vect{C},\vect{B}_j^{b},\vect{A}_i^{a}
\end{array}\right.
$$

\noindent Hence:

$$(A\comp{i} (B\comp{j} C))^\sharp = ((A\comp{i} B)\comp{i+j-1} C)^\sharp$$

\noindent For the case $(A\comp{i} B)\comp{k} C$, if one puts $k=i+j-1$ one gets $j=k-i+1$. Therefore, changing indices: we have that:

$$((A\comp{i} B)\comp{j} C)^\sharp = (A\comp{i} (B\comp{j-i+1} C))^\sharp$$

This ends the case of discontinuous associativity.

\item Mixed permutation:\\

\noindent There are two cases: $B\prec_A C$ or $C\prec_A B$. We consider only the first case, i.e.\ $B\prec_A C$. The
other case is analogous. Let us see $((A\comp{i} B)\comp{j} C)^\sharp$:

$$\vect{A}|_i\vect{B} = \vect{A}_0^{i-1},\vect{B},\vect{A}_i^{k-1},
\underbrace{\sep}_{j\scriptsize{\mbox{-th }\sep}},\vect{A}_k^{a}$$

\noindent We have therefore:

$$j=k-1+b\mbox{ iff }k=j-b+1$$

$$((A\comp{i}B)\comp{j} C)^\sharp = \vect{A}_0^{i-1},\vect{B},\vect{A}_i^{k-1},
\vect{C},\vect{A}_k^{a}$$

\noindent Hence:

$$(\vect{A}\comp{j-b+1} \vect{C})^\sharp = \vect{A}_0^{i-1},\sep,\vect{A}_i^{k-1},
\vect{C},\vect{A}_k^{a}$$

It follows that:

$$((A\comp{j-b+1}C)\comp{i} B)^\sharp= \vect{A}_0^{i-1},\vect{B},\vect{A}_i^{k-1},
\vect{C},\vect{A}_k^{a}$$

\noindent Summarizing:

$$
\left \{
\begin{array}{l}
((A\comp{i} B)\comp{j} C)^\sharp = \vect{A}_0^{i-1},\vect{B},\vect{A}_i^{k-1},
\vect{C},\vect{A}_k^{a}\\
((A\comp{j-b+1}C)\comp{i} B)^\sharp= \vect{A}_0^{i-1},\vect{B},\vect{A}_i^{k-1},
\vect{C},\vect{A}_k^{a}
\end{array} \right.
$$

\noindent Hence
$$
\begin{array}{l}
((A\comp{i} B)\comp{j} C)^\sharp = ((A\comp{j-b+1}C)\comp{i} B)^\sharp
\end{array}
$$

\noindent Putting $i= j-b+1$ we have that $j=i+b-1$. Hence:

$$((A\comp{i} C)\comp{j} B)^\sharp = ((A\comp{j}C)\comp{i+b-1} B)^\sharp$$

\noindent This ends the case of mixed permutation.

\item Split-wrap:\\
We have:

$$
\begin{array}{lll}
 ((A\comp{}\mathbb{J})\comp{a+1} B)^\sharp &=& (\vect{A},\sep)|_{a+1}\vect{B}=\vect{A},\vect{B}\\
 ((\mathbb{J}\comp{} B)\comp{1} A)^\sharp &=& (\sep,\vect{B})|_1\vect{A}=\vect{A},\vect{B}
\end{array}
$$
\noindent Hence:

$$
\begin{array}{lll}
((A\comp{}\mathbb{J})\comp{a+1} B)^\sharp &=& (A\comp{} B)^\sharp\\
&\mbox{ and }&\\
((\mathbb{J}\comp{} B)\comp{1} A)^\sharp &=& (A\comp{} B)^\sharp
\end{array}
$$

\noindent This ends the case of split-wrap.
\item Units:

$$
\begin{array}{lllll}
(\mathbb{I}\comp{} A)^\sharp&=&\vect{A}&=& (A\comp{}\mathbb{I})^\sharp\\
(\mathbb{J}\comp{1}A)^\sharp&=&(\sep|_1\vect{A})&=&\vect{A}=\vect{A}|_i\sep=(A\comp{i}\mathbb{J})^\sharp
\end{array}
$$

We recall that types play the role of variables of structural terms.
Now, we have seen that structural rules for arbitrary type variables collapse into the same textual
form. This result generalizes to arbitrary structural terms by simply using type substitution. 

More concretely, we have proved that: if $T\equi S$ (i.e.\ $S$ is the result of applying a single structural rule to $T$)
then $T^\sharp= S^\sharp$. Suppose we have $T\equi^* S$ (we omit the trivial case $T\equi^* T$). We have then a chain:

$$T:=T_1\equi T_2\equi\cdots \equi T_{i-1}\equi T_i=:S\mbox{ for }i\geq 2$$

\noindent Applying $(\cdot)^\sharp$ to each $T_k\equi T_{k+1}$ $(1\leq k\leq i-1)$ we have proved that:

$$(T_k)^\sharp = (T_{k+1})^\sharp$$

\noindent We have therefore a chain of identities:

$$(T)^\sharp=(T_1)^\sharp= (T_2)^\sharp=\ldots=(T_i)^\sharp=(S)^\sharp$$

\noindent This completes the proof.
\end{itemize}

% End collapsing structural rules
}% end proof

\noindent We will now prove the associativity theorems of $\mathbf{hD}$ displayed in (\ref{chapt4disphdthms}).
Other theorems corresponding to the structural postulates of $\mathbf{mD}$ have similar proofs.

\begin{itemize}
 \item Continuous associativity is obvious as in the Lambek calculus. The only difference
is that types are sorted and in our notation the antecedent of hypersequents have the vectorial
notation.  
 \item Discontinuous associativity: we suppose that $B\between_A C$. The following hypersequents
are provable:

$$
\vect{(A\odot_i B)\odot_{i+j-1}C}  \yields  A\odot_i(B\odot_j C)
$$

\noindent And:

$$
\vect{A\odot_i(B\odot_j C)}  \yields (A\odot_i B)\odot_{i+j-1}C 
$$

By the previous lemma the identity $\vect{A}|_i(\vect{B}|_j \vect{C})= (\vect{A}|_i\vect{B})|_{i+j-1} \vect{C}$ holds.
We have the two following hypersequent derivations:

$$
\prooftree
\prooftree
\prooftree
\prooftree
\vect{A}\yields A
\tb
\prooftree
\vect{B}\yields B
\tb
\vect{C}\yields C
\justifies
\vect{B}|_j\vect{C}\yields B\odot_j C
\using
\odot_j R
\endprooftree
\justifies
\vect{A}|_i(\vect{B}|_{j}\vect{C})=(\vect{A}|_i\vect{B})|_{i+j-1} \vect{C}\yields A\odot_i(B\odot_j C)
\using \odot_i R 
\endprooftree
\justifies
(\vect{A}|_i\vect{B})|_{i+j-1} \vect{C}
\yields A\odot_i(B\odot_j C)
\using
\endprooftree
\justifies
\vect{A\odot_iB}|_{i+j-1} \vect{C}
\yields A\odot_i(B\odot_jC)
\using \odot_i L
\endprooftree
\justifies
\vect{(A\odot_iB)\odot_{i+j-1} C}
\yields A\odot_i(B\odot_jC)
\using\odot_{i+j-1} L
\endprooftree
$$

\noindent \hspace{5cm}and

$$
\prooftree
\prooftree
\prooftree
\prooftree
\vect{A}\yields A \tb \vect{B}\yields B
\justifies
\vect{A}|_i\vect{B}\yields (A\odot_i B)
\using
\odot_{i}R
\endprooftree
\vect{C}\yields C
\justifies
(\vect{A}|_i\vect{B})|_{i+j-1} \vect{C}=\vect{A}|_i(\vect{B}|_j \vect{C})\yields (A\odot_iB)\odot_{i+j-1}C
\using
\odot_{i+j-1}R
\endprooftree
\justifies
\vect{A}|_i(\vect{B\odot_j C})\yields (A\odot_iB)\odot_{i+j-1}C
\using \odot_j L
\endprooftree
\justifies
\vect{A\odot_i(B\odot_j C)}\yields 
(A\odot_iB)\odot_{i+j-1}C
\using\odot_i L
\endprooftree
$$

\end{itemize}

\thm{Equivalence Theorem for $\mathbf{StructTerm}$}
{
Let $R$ and $S$ be arbitrary structural terms. The following holds:
$$R\equi^* S\mbox{ iff } (R)^\sharp = (S)^\sharp$$
}\label{chapt4equivthmstructerm}% end theorem normal form \mathbf{structTerm}
\prf{
We have already seen in (\ref{chapt4hDabsoptionSR}) the \emph{only if} case, which is the fact that $\mathbf{hD}$ absorbs
the $\mathbf{Eq_D}$ structural rules.
The \emph{if} case is more difficult and needs some technical machinery from sorted universal algebra. For details, see
\cite{valentin:phd}.
}

\lema{$(\cdot)^\sharp$ is an Epimorphism}{
\noindent For every $\Delta\in\mathcal{O}$ there exists a structural term\footnote{In fact there 
exists an infinite set of such structural terms.} $T_\Delta$ such that:

$$(T_\Delta)^\sharp = \Delta$$
}\label{chapt4sharpmapisEpi}% end lemma
\prf{
\noindent This can be proved by induction on the structure of hyperconfigurations.
We define recursively $T_\Delta$ such that $(T_\Delta)^\sharp= \Delta$:

\begin{itemize}
 \item Case $\Delta=\Lambda$ (the empty tree): $T_\Delta=\mathbb{I}$.
 \item Case where $\Delta= A,\Gamma$: $T_\Delta = A\comp{} T_\Gamma$, where by induction hypothesis (i.h.) 
$(T_\Gamma)^\sharp = \Gamma$.
\item Case where $\Delta= \sep,\Gamma$: $T_\Delta = \mathbb{J}\comp{} T_\Gamma$, where by i.h.\ 
$(T_\Gamma)^\sharp = \Gamma$.
\item Case $\Delta=\vect{A}\otimes\langle\Delta_1,\cdots,\Delta_a\rangle,\Delta_{a+1}$. By i.h.\ we have:

$$(T_{\Delta_i})^\sharp = \Delta_i\mbox{ for }1\leq i\leq a+1$$
$$
\begin{array}{l}
 T_\Delta=(A\comp{1}T_{\Delta_1})\comp{}T_{\Delta_2}\mbox{ if }a=1\\
 T_\Delta=((\cdots((A\comp{1}T_{\Delta_1})\comp{1+d_1}T_{\Delta_2})\cdots)\comp{1+d_1+\cdots+d_{a-1}} T_{\Delta_{a}})\comp{}T_{\Delta_{a+1}}\mbox{ if }a>1
\end{array}
$$

\end{itemize}
}\\

By induction on the structure of \structterm{}, we have the following intuitive result on the
relationship between structural contexts and hypercontexts:

\disp{
$(T[S])^\sharp= T^\sharp\langle S^\sharp\rangle$
}\label{chapt4relstructcontexthypercontexts}

\noindent These two technical results we have seen above are necessary for the proof of the faithful embedding translation $(\cdot)^\sharp$ of theorem (\ref{chapt4sharpembeddingthm}). We prove now an important theorem which is crucial for the mentioned theorem (\ref{chapt4sharpembeddingthm}).

\thm{Visibility for Extraction in $\mathbf{StructTerm}$}{
Let $T[A]$ be a structural term with a distinguished occurrence of type $A$.
Suppose that:

$$(T[A])^\sharp =\Delta|_i \vect A$$

where $\Delta\in\mathcal{O}$ and $A\in\mathcal{F}$. Then $A$ is visible for extraction in $T[A]$, i.e. there exist a structural term $T'$ and an index $i$ such that:

$$T[A]\,\equi^*\, T'\comp{i}A$$

}\label{chapt4visibility4extraction}% end theorem visibility for extraction
\prf{
Let $T_\Delta$ be a structural term such that $(T_\Delta)^\sharp=\Delta$. This is possible by lemma (\ref{chapt4sharpmapisEpi}).
We have $(T_\Delta\comp{i} A)^\sharp= \Delta|_i\vect{A}$. We have then $(T_\Delta\comp{i} A)^\sharp= (T[A])^\sharp$. 
By the equivalence theorem (\ref{chapt4equivthmstructerm}) it follows that $T[A]\equi^*T_\Delta\comp{i} A$.
Put $T':=T_\Delta$. We are done. 
}\label{extractability}\\% end proof theorem visibility for extraction

\thm{Uniqueness of Extractability}
{Suppose that $T[A]\sim S\comp{i} A$ and $T[A]\sim S'\comp{j} A$, where $A$. Then:
$$
\begin{array}{lcl}
S&\sim^*& S'\\
i&=&j
\end{array}
$$
}\label{uniquenessextractability}
\prf{
We have that $(S\comp{i} A)^\sharp=\Delta|_i\vect A=\Delta|_j\vect A=
(S'\comp{j} A)^\sharp$. Hence $i=j$ and $(S)^\sharp = (S')^\sharp$. By theorem
(\ref{chapt4equivthmstructerm}), $S\sim^* S'$. We are done.
}% end proof on uniqueness of extractability

\noindent Theorems (\ref{extractability}) and (\ref{uniquenessextractability}) will be crucial for the proof
of the $(\cdot)^\sharp$ embedding theorem (\ref{chapt4sharpembeddingthm}). 

\noindent Before proving theorem (\ref{chapt4sharpembeddingthm}), it is worth seeing what is the intuition
behind the structural rules of \Eqd{}. This intuition is exemplified by a constructive proof of theorem
(\ref{extractability}):\\
\prf{Constructive proof of theorem (\ref{extractability}):
% insertion visibility for extraction Chapter 2
By induction on the structural complexity of $T[A]$: 
The cases are as follows:
\begin{itemize}
 \item[i)] $T[A]= A$.\\

We put $T'= \mathbb{J}$ and hence :

\begin{center}
$T[A]\sim^* \mathbb{J}\comp{1} A$
\end{center}
\item[ii)] $T[A]= S[A]\comp{} R$.\\

By induction hypothesis (i.h.), $S[A]\sim^* S'\comp{k} A$ for some $k>0$. We have the following
equational derivation:

\begin{flushleft}
$
\begin{array}{llll}
T[A]&\sim^*&(S'\comp{k} A)\comp{} R&\\
&\sim^*&(\mathbb{J}\comp{} R)\comp{1}(S'\comp{k} A) &\mbox{ by }\mathbf{SW}\\
&\sim^*&((\mathbb{J}\comp{} R)\comp{1} S')\comp{k} A) &\mbox{ by }\mathbf{Assc_d}\\
&\sim^*&(S'\comp{} R)\comp{k} A &\mbox{ by }\mathbf{SW}

\end{array}
$
\end{flushleft}

%\noindent In tree format:
%\begin{flushleft}
%\qroofx=1
%\qroofy=1
%$
%\begin{array}{lll}
%\Tree [.$\oplus$ \qroof{$\cdots A\cdots $}.$S[A]$  $R$ ]&\leadsto&\Tree [.$\otimes_i$ [.$\oplus$ $S'$  $R$ ] $A$ ]
%\end{array}
%$ 
%\end{flushleft}
 \item[iii)] $T[A]= S\comp{} R[A]$\\

By i.h.\ $R[A]\sim^* R'\comp{k} A$ for some term $R'$ and $k>0$. It follows that:

\begin{flushleft}
$
\begin{array}{llll}
 T[A]&\sim^*& S\comp{}(R'\comp{k} A)&\\
&\sim^*&(S\comp{} \mathbb{J})\comp{S(S)+1}(R'\comp{k} A) &\mbox{ by }\mathbf{SW}\\
&\sim^*&((S\comp{} \mathbb{J})\comp{S(S)+1}R')\comp{S(S)+k} A &\mbox{ by }\mathbf{Assc_d}\\
&\sim^*&(S\comp{} R')\comp{S(S)+k} A &\mbox{ by }\mathbf{SW}

\end{array}
$
\end{flushleft}

%\noindent In tree format:
%\begin{flushleft}
%\qroofx=1
%\qroofy=1
%$
%\begin{array}{lll}
%\Tree [.$\oplus$  $S$ \qroof{$\cdots A\cdots $}.$R[A]$ ]&\leadsto&\Tree [.$\otimes_{S(x)+k}$ [.$\oplus$ $S$  $R'$ ] $A$ ]
%\end{array}
%$ 
%\end{flushleft}
 \item[iv)] $T[A]= S[A]\comp{i} R$ for some term $S[A]$, $R$ and $i>0$.\\

By i.h.\ $S[A]\sim^* S'\comp{k} A$ for some $S'$ and $i>0$. We derive the
following equation:

\begin{flushleft}
$
\begin{array}{llll}
 T[A]&\sim^*& (S'\comp{k} A)\comp{i} R &
\end{array}
$
\end{flushleft}

If $R=\mathbb{J}$ we are done. Suppose that $R\neq\mathbb{J}$. In this case $A$ must permute with $R$ in $S'$, i.e.\ $A\prec_{S'}R$ or $R\prec_{S'}A$, for otherwise $(T[A])^\sharp=\Delta|_i\vect{A}$ would not hold.
Without loss of generality, let us suppose that $A\prec_{S'}R$. In that case we have:

\begin{flushleft}
$
\begin{array}{llll}
 T[A]&\sim^*& (S'\comp{i-S(A)+1} R)\comp{k} A&\mbox{by }\mathbf{MixPerm1}
\end{array}
$
\end{flushleft}

\noindent Hence $A$ is permutated to right periphery in $T[A]$.
%\begin{flushleft}
%\qroofx=1
%\qroofy=1
%$
%\begin{array}{lll}
%\Tree [.$\otimes_i$  \qroof{$\cdots A\cdots $}.$S[A]$ $R$  ]&\leadsto&\Tree [.$\otimes_{k}$ [.$\otimes_{i-S(x)+1}$ $S'$  $R$ ] $A$ ]
%\end{array}
%$ 
%\end{flushleft}
 \item[v)] $T[A]=S\comp{i} R[A]$ for some terms $S$ and  $R[A]$ and $i>0$. By i.h.\ \\
$R[A]\sim^* R'\comp{k} A$. Then:

\begin{flushleft}
$
\begin{array}{llll}
 T[A]&\sim^*& S\comp{i} (R'\comp{k} A) &\\
     &\sim^*& (S\comp{i} R')\comp{i+k-1}A&\mbox{by }\mathbf{Assc_d}
\end{array}
$
\end{flushleft}

%\noindent In tree format:\\
%
%\begin{flushleft}
%\qroofx=1
%\qroofy=1
%$
%\begin{array}{lll}
%\Tree [.$\otimes_i$   $S$ \qroof{$\cdots A\cdots $}.$R[A]$  ]&\leadsto&\Tree [.$\otimes_{i+k-1}$ [.$\otimes_{i}$ $S'$  $R$ ] $A$ ]
%\end{array}
%$ 
%\end{flushleft}

\end{itemize}

% end insertion

}
\rmark{Interestingly, the constructive proof for extractability does not use continuous associativity.
Therefore, a priori a non-associative discontinuous Lambek calculus could be considered. This remark
needs further study. 
}

%%%%%%%%%%%%%%%%%%%%%%%%%%%%%%%%%%%%%%%%%%%%%%%%%%%%%%%
% (.)^\sharp embedding theorem
%%%%%%%%%%%%%%%%%%%%%%%%%%%%%%%%%%%%%%%%%%%%%%%%%%%%%%%

\thm{Faithfulness of $(\cdot)^\sharp$ Embedding Translation}{
Let $A,\mbox{ }X$ and $\Delta$ be respectively a type, a structural term and a hyperconfiguration.
The following statements hold:

\begin{itemize}
 \item[i)] $\mbox{If }\vdash_{\mathbf{mD}} X\oneyields A\:
\mbox{then}\:\vdash_{\mathbf{hD}} (X)^\sharp\yields A$

\item[ii)] $\mbox{For any }X\mbox{ such that }(X)^\sharp=\Delta,\mbox{ if }\vdash_{\mathbf{hD}}\Delta\yields A
\mbox{ then }\vdash_{\mathbf{mD}}X\oneyields A$
\end{itemize}
}\label{chapt4sharpembeddingthm}
\prf{\begin{itemize}
 \item[i)] Logical rules in $\mathbf{mD}$ translate without any problem to $\mathbf{hD}$. We need recall only that
if $X$ and $Y$ are structural terms then $(X\comp{} Y)^\sharp= (X)^\sharp, (Y)^\sharp$ and $(X\comp{i} Y)^\sharp
=(X)^\sharp|_i(Y)^\sharp$. Structural rules in $\mathbf{mD}$ collapse in the same textual form as theorem
(\ref{chapt4hDabsoptionSR}) proves. Finally, the Cut rule has no surprise. This proves i).

\item[ii)] This part of the theorem becomes easy if we use the following four facts:

\begin{itemize}
 \item Lemma (\ref{chapt4sharpmapisEpi}) which states that for any hyperconfiguration $\Delta$ there is a structural
term $T_\Delta$ such that $(T_\Delta)^\sharp = \Delta$.

 \item The fact (\ref{chapt4relstructcontexthypercontexts}) we stated before which gives the relationship between structural terms and
hypercontexts $(T[A])^\sharp = T^\sharp\langle\vect{A}\rangle$.
 
 \item Theorem (\ref{chapt4equivthmstructerm}).

 \item Theorem (\ref{chapt4visibility4extraction}).
\end{itemize}

The proof is by induction on the length of $\mathbf{hD}$ derivations. The three first facts prove the induction of all the rules
but the right rule of the connectives $\extract_i$. Suppose the last rule of a $\mathbf{hD}$ derivation is $\extract_i R$:

$$
\prooftree
\Delta|_i\vect{A}\yields B
\justifies
\Delta\yields B\extract_i A
\using\extract_i R
\endprooftree
$$
Let $T[A]$ be such that $(T[A])^\sharp=\Delta|_i\vect{A}$. We know by induction hypothesis that $\vdash_{\mathbf{mD}}T[A]\yields B$.
By the last fact of above, i.e.\ theorem (\ref{chapt4visibility4extraction}) of visibility
of extraction, since $(T[A])^\sharp=\Delta|_i\vect{A}$, we know there exist $T'$ and $i$ such that $T[A]\equi^* T'\comp{i}A$.
It follows that in $\mathbf{mD}$:

$$
\prooftree
\prooftree
\prooftree
T[A]\oneyields B
\justifies
\hspace{4cm}\vdots\hspace{0.5cm}\mbox{\textbf{Sequence of structural rules}}
\endprooftree
\justifies
T'\comp{i}A\oneyields B
\endprooftree
\justifies
T'\oneyields B\extract_i A
\using\extract_i R
\endprooftree
$$

Hence, $\vdash_{\mathbf{mD}}T'\oneyields B\extract_i A$. And $T'$ is in fact $T_\Delta$, and therefore $(T')^\sharp=\Delta$.
Moreover, for any $S$ such that $(S)^\sharp\equi^* T'$, we have that applying a finite number of structural rules we obtain 
the $\mathbf{mD}$ provable sequent $S\oneyields B\extract_i A$, and of course $(S)^\sharp=\Delta$. This completes the proof of ii). 

\end{itemize}

}% End of the proof of the (.)^\sharp embedding theorem

\examplewt{
Let $B,D, E, C,A$ five arbitrary atomic types of sort respectively $2$, $2$, $1$, $0$ and $0$.
The following two derivations have the following end-sequent and end-hypersequent:
$$
\begin{array}{l}
\vdash_{\mathbf{mD}}(((B\extract_2 A\comp{1}D)\comp 4 E)\comp 3 (\mathbb{J}\comp{} C\bsl A))\oneyields ((B\odot_1 D)\odot_3E)\extract_3 C\\
\vdash_{\mathbf{hD}}\seg{0}{B\extract_2 A},\vect{D},\seg{1}{B\extract_2 A},[],C\bsl A,\seg{2}{B\extract_2 A},\vect{E},\seg{3}{B\extract_2 A}\yields ((B\odot_1 D)\odot_3 E)\extract_3 C
\end{array}
$$
\noindent The above multimodal sequents are in correspondence through the mapping $(\cdot)^\sharp$.
Derivations (\ref{mmder}) and (\ref{hsder}) illustrate theorem (\ref{chapt4sharpembeddingthm}).
Notice the sequence of structural rules in derivation (\ref{mmder}) in order to extract type $C$.
\vspace{0.5cm}
\disp{
$
\prooftree
\prooftree
\prooftree
\prooftree
\prooftree
\prooftree
\prooftree
\prooftree
C\comp{} C\bsl A\oneyields A
\tb
B\extract_2 A\comp{2}A\oneyields B
\justifies
B\extract_2 A\comp{2}(C\comp{} C\bsl A)\oneyields B
\using\extract_2
\endprooftree
\tb D\oneyields D
\justifies
(B\extract_2 A\comp{2}(C\comp{} C\bsl A))\comp{1}D\oneyields B\odot_1 D
\using\odot_1
\endprooftree
\tb E\oneyields E
\justifies
((B\extract_2 A\comp{2}(C\comp{} C\bsl A))\comp{1}D)\comp{3}E\oneyields (B\odot_1 D)\odot_3E
\using\odot_3
\endprooftree
\justifies
((B\extract_2 A\comp{1}D)\comp{3}(C\comp{} C\bsl A))\comp 3 E\oneyields (B\odot_1 D)\odot_3E
\using \mathbf{MixPerm}
\endprooftree
\justifies
((B\extract_2 A\comp{1}D)\comp 4 E)\comp 3 (C\comp{} C\bsl A)\oneyields (B\odot_1 D)\odot_3E
\using \mathbf{MixPerm}
\endprooftree
\justifies
((B\extract_2 A\comp{1}D)\comp 4 E)\comp 3 ((\mathbb{J}\comp{} C\bsl A)\comp 1 C)\oneyields (B\odot_1 D)\odot_3E
\using \mathbf{SW}
\endprooftree
\justifies
(((B\extract_2 A\comp{1}D)\comp 4 E)\comp 3 (\mathbb{J}\comp{} C\bsl A))\comp 3 C\oneyields (B\odot_1 D)\odot_3E
\using \mathbf{Assc_d}
\endprooftree
\justifies
(((B\extract_2 A\comp{1}D)\comp 4 E)\comp 3 (\mathbb{J}\comp{} C\bsl A))\oneyields ((B\odot_1 D)\odot_3E)\extract_3 C
\using\extract_3
\endprooftree
$
}\label{mmder}
}
\vspace{0.5cm}
\disp{
{\normalsize%\small 
$
\prooftree
\prooftree
\prooftree
\prooftree
C, C\bsl A\yields A
\tb
\seg{0}{B\extract_2 A},\sep,\seg{1}{B\extract_2 A},A,\seg{2}{B\extract_2 A},\sep,\seg{3}{B\extract_2 A}\yields B
\justifies
\seg{0}{B\extract_2 A},\sep,\seg{1}{B\extract_2 A},C,C\bsl A,\seg{2}{B\extract_2 A},\sep,\seg{3}{B\extract_2 A}\yields B
\using\extract_2
\endprooftree
\tb \vect{D}\yields D
\justifies
\seg{0}{B\extract_2 A},\vect{D},\seg{1}{B\extract_2 A},C,C\bsl A,\seg{2}{B\extract_2 A},\sep,\seg{3}{B\extract_2 A}\yields B\odot_1 D
\using\odot_1
\endprooftree
\tb E\yields E
\justifies
\seg{0}{B\extract_2 A},\vect{D},\seg{1}{B\extract_2 A},C,C\bsl A,\seg{2}{B\extract_2 A},\vect{E},\seg{3}{B\extract_2 A}\yields (B\odot_1 D)\odot_3 E
\using\odot_3
\endprooftree
\justifies
\seg{0}{B\extract_2 A},\vect{D},\seg{1}{B\extract_2 A},[],C\bsl A,\seg{2}{B\extract_2 A},\vect{E},\seg{3}{B\extract_2 A}\yields ((B\odot_1 D)\odot_3 E)\extract_3 C
\using \extract_3
\endprooftree
$}
}\label{hsder}

%\begin{flushleft}
%$
%\begin{array}{lll}
%((B\extract_2 A\comp{2}(C\comp{} C\bsl A))\comp{1}D)\comp{3}E&\stackrel{\mathbf{MixPerm}}{\sim}&\\
%((B\extract_2 A\comp{1}D)\comp{3}(C\comp{} C\bsl A))\comp 3 E&\stackrel{\mathbf{MixPerm}}{\sim}&\\
%((B\extract_2 A\comp{1}D)\comp 4 E)\comp 3 (C\comp{} C\bsl A)&\stackrel{\mathbf{SW}}{\sim}&\\
%((B\extract_2 A\comp{1}D)\comp 4 E)\comp 3 ((\mathbb{J}\comp{} C\bsl A)\comp 1 C)&\stackrel{\mathbf{Assc_d}}{\sim}&\\
%(((B\extract_2 A\comp{1}D)\comp 4 E)\comp 3 (\mathbb{J}\comp{} C\bsl A))\comp 3 C
%\end{array}
%$
%\end{flushleft}

\section{Conclusions}
It is not a priori a trivial task to find out a set of structural rules $\mathcal{E}$ that makes the hypersequent calculus
$\mathbf{hD}$ equivalent to a multimodal calculus with the structural rules of $\mathcal{E}$.
The faithful embedding translation $(\cdot)^\sharp$ between $\mathbf{mD}$ and $\mathbf{hD}$ is then, we think, a remarkable discovery. 
The equivalent multimodal calculus $\mathbf{mD}$ gives $\mathbf{D}$ the Moot's powerful proof net 
machinery almost for free (see \cite{DBLP:journals/corr/abs-0711-2444}). 
It must be noticed that this proof net theory approach for $\mathbf{D}$ is completely different from the one in \cite{morfad:lac08}.
Finally, the discovery of $\mathbf{mD}$ can be very useful to investigate new soundness and completeness results
for $\mathbf{D}$ (see \cite{valentin:phd}).
\bibliographystyle{plain}
\bibliography{bib_lambek_hidden}%{bib_dlc}

\end{document}